\documentclass[aps,prl,reprint,superscriptaddress,showpacs]{revtex4-1}

\usepackage{graphicx}
\usepackage{dcolumn}
\usepackage{bm}
\usepackage{hyperref}

\usepackage{float}
\usepackage[separate-uncertainty=true]{siunitx}
\usepackage{braket}
\usepackage{mhchem}
\usepackage[utf8]{inputenc}
\usepackage[T1]{fontenc}

\graphicspath{{figures/}}

\begin{document}

\preprint{AIP/123-QED}
\title{Parity switching in a full-shell superconductor-semiconductor nanowire qubit}%

\newcommand{\qdev}{Center for Quantum Devices, Niels Bohr Institute, University of Copenhagen, 2100 Copenhagen, Denmark}

\newcommand{\eth}{Laboratory for Solid State Physics, ETH Z\"urich, CH-8093 Z\"urich, Switzerland}

\author{O.~Erlandsson}
\thanks{These authors contributed equally to this work}
\affiliation{\qdev}

\author{D.~Sabonis}
\thanks{These authors contributed equally to this work}
\affiliation{\qdev}
\affiliation{\eth}

\author{A.~Kringhøj}
\affiliation{\qdev}

\author{T.W.~Larsen}
\affiliation{\qdev}

\author{P.~Krogstrup}
\affiliation{\qdev}

\author{K.~D.~Petersson}
\affiliation{\qdev}

\author{C.~M.~Marcus}
\affiliation{\qdev}

\date{\today}

\begin{abstract}

The rate of charge-parity switching in a full-shell superconductor-semiconductor nanowire qubit is measured by directly monitoring the dispersive shift of a readout resonator. At zero magnetic field, the measured switching time scale ${ T_P }$ is on the order of ${ \SI{100}{\milli\second} }$. Two-tone spectroscopy data post-selected on charge-parity is demonstrated. With increasing temperature or magnetic field, ${ T_P }$ is at first constant, then exponentially suppressed, consistent with a model that includes both non-equilibrium and thermally activated quasiparticles. As ${ T_P }$ is suppressed, qubit lifetime ${ T_1 }$ also decreases. The long ${ T_P \sim \SI{0.1}{\second}}$ at zero field is promising for future development of qubits based on hybrid nanowires.
\end{abstract}

	\maketitle
	
The coherent manipulation of any transmon qubit is threatened by the presence of unpaired quasiparticles (QPs) in the superconductor, a problem referred to as quasiparticle poisoning (QPP). Understanding the nature and rate of QPP is therefore part of the challenge of using these superconducting circuits to construct complex quantum information processing devices. In a transmon qubit \cite{kochChargeinsensitiveQubitDesign2007}, a Josephson junction (JJ) with an associated energy ${ E_J }$ separates two superconductors which are shunted by a large capacitance with a charging energy ${ E_C }$. As QPs tunnel across the JJ, they contribute to relaxation if they carry energy matching the qubit transition, and to dephasing in the case of non-vanishing charge dispersion.

Hybrid superconductor-semiconductor nanowire (NW) variants of the transmon qubit \cite{delangeRealizationMicrowaveQuantum2015, larsenSemiconductorNanowireBasedSuperconductingQubit2015} provide an alternative to conventional metallic systems. The semiconductor platform offers advantages such as control of ${ E_J }$ by electrostatic gating, and reduced charge dispersion without loss of anharmonicity \cite{bargerbosObservationVanishingCharge2020, kringhojSuppressedChargeDispersion2020}. The system has been developed to be compatible with large magnetic fields \cite{luthiEvolutionNanowireTransmon2018, krollMagneticFieldResilientSuperconductingCoplanarWaveguide2019, kringhojMagneticFieldCompatibleSuperconductingTransmon2021, uilhoornQuasiparticleTrappingOrbital2021} where it potentially could be used for topological quantum computation \cite{ginossar2014microwave, vaitiekenas2020flux}. With a magnetic field applied, NWs with a fully covering superconducting shell have been shown to exhibit destructive Little-Parks effect, both in dc transport and qubit measurements \cite{vaitiekenasAnomalousMetallicPhase2020, sabonisDestructiveLittleParksEffect2020}.

 \begin{figure}[!th]
\includegraphics[width=8cm]{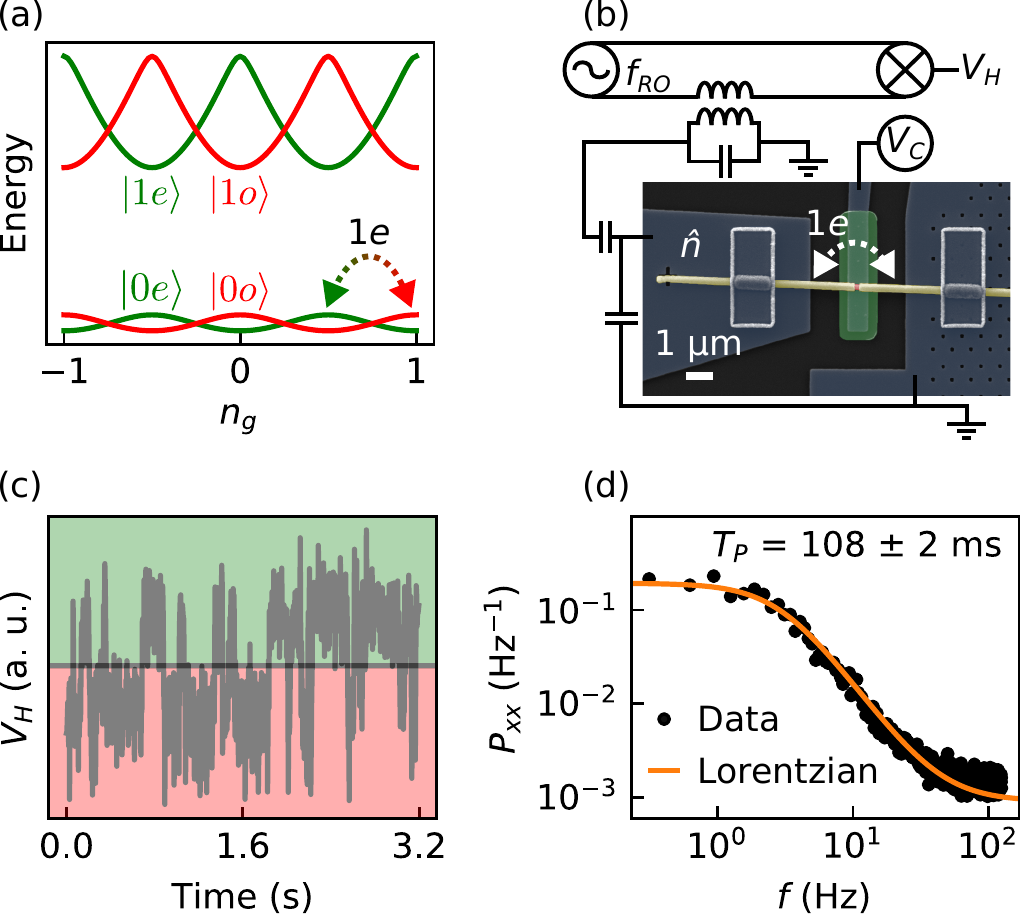}
\caption{\label{fig:fig1} (a) Parity branches of ground and excited states of the transmon Hamiltonian. The dotted arrow symbolizes a charge-parity switch (poisoning) as a quasiparticle tunnels across the Josephson junction. (b) False-color scanning electron micrograph of device and diagram of surrounding cQED circuit. An InAs (red) nanowire with an Al (yellow) shell is embedded in a \ce{NbN} (blue; rectangular contacts to nanowire are \ce{NbTiN}) structure on a Si substrate (black). A layer of \ce{HfO2} (green) separates the bottom gate ${ V_C }$ from the etched junction. (c) Time trace of demodulated transmission ${ V_H }$ through the readout resonator, showing discrete switching, interpreted as changes of charge-parity state. Data taken at ${ T = \SI{30}{\milli\kelvin} }$ and $B=0$. The horizontal line between differently colored regions indicates data binning threshold. (d) Average power spectrum density of 40 traces fitted to a Lorentzian distribution, with ${ T_P }$ as given by the fit.}
\end{figure}
 
 In this Letter, the rate of QPP is measured in a full-shell superconductor-semiconductor hybrid NW qubit. It is found that QPP occurs on a timescale of ${ T_P \sim 0.1\,}$s, which is far from limiting the coherence time of the qubit. As either the temperature or magnetic field is increased, ${ T_P }$ is first constant then exponentially suppressed, suggesting the existence of both a non-equilibrium and equilibrium population of QPs in the qubit. Additionally, while the qubit frequency recovers in the first Little-Parks lobe, ${ T_P }$ remains below our detectable range (not sufficiently exceeding ${ \SI{30}{\micro\second} }$) after the initial suppression.
 
QPP in transmons was first measured directly by a Ramsey-type pulse sequences sensitive to the charge-parity state of the qubit \cite{risteMillisecondChargeparityFluctuations2013}. This technique was subsequently used to study transitions across all four parity-logical states \cite{serniakHotNonequilibriumQuasiparticles2018} and the effect of shielding and qubit geometry on QPP levels \cite{kurter2021quasiparticle, gordon2021environmental}. A technique based on direct dispersive monitoring of the charge-parity has also been developed \cite{serniakDirectDispersiveMonitoring2019}. This was recently used to study QPP in NWs similar to those investigated here but with the superconducting shell only covering half the facets, where it was found that QPP occured on the timescale of \SI{100}{\micro\second} and could have a non-monotonic dependence on magnetic field \cite{uilhoornQuasiparticleTrappingOrbital2021}. The effect of geometry was also studied with the dispersive technique \cite{pan2022engineering}. Typical reported QPP time scales have ranged from \SI{100}{\micro\second} up to \SI{1}{\second} \cite{risteMillisecondChargeparityFluctuations2013, serniakHotNonequilibriumQuasiparticles2018, serniakDirectDispersiveMonitoring2019, uilhoornQuasiparticleTrappingOrbital2021, gordon2021environmental, kurter2021quasiparticle, pan2022engineering}. Methods to mitigate unwanted quasiparticles in superconducting circuits have included qubit pumping \cite{gustavssonSuppressingRelaxationSuperconducting2016} and trapping of quasiparticles by vortices \cite{voolNonPoissonianQuantumJumps2014, wangMeasurementControlQuasiparticle2014} or regions of normal metal \cite{patelPhononMediatedQuasiparticlePoisoning2017}. 

QPP is measured as the rate of transitions between the two parity branches of the transmon. How QPP leads to dephasing (for non-vanishing charge dispersion) can be understood by considering the transmon Hamiltonian, ${ \hat H = 4 E_C (\hat n - n_g)^2 - E_J \cos \hat \phi }$, where ${ n }$ is the number of Cooper pairs on the island, ${ n_g }$ is the the effective offset charge and ${ \phi }$ is the phase difference across the JJ. As indicated in Fig.~\ref{fig:fig1}(a), each time a poisoning event occurs ${ n_g }$ is shifted by ${ 1e }$, which except for at degeneracy points, changes the qubit transition frequency, contributing to qubit dephasing and establishing two distinct charge-parity branches (even and odd) of each logical state. The charge-parity state is measured by monitoring the microwave transmission through the resonator, sensitive to changes in qubit transition frequency via the dispersive shift.


We now describe the experimental set-up to realize a transmon using a NW. To fabricate the sample, a 35~nm thick \ce{NbN} film was sputtered onto a \ce{Si} substrate. The cQED architecture, consisting of a microwave feedline, readout resonators, qubit islands, gate lines, and ground plane, was patterned in the \ce{NbN} film by reactive ion etching. The cQED circuit is shown in Fig.~\ref{fig:fig1}(b). For the qubit studied here, the island was designed for a charging energy ${ E_C/h = \SI{500}{\mega\hertz} }$ and the resonator frequency ${ f_{\textrm{RO}} = \SI{5.4}{\giga\hertz} }$. The islands were connected to the ground plane by micromanipulator deposited VLS-grown \ce{InAs} NWs with a full superconducting \ce{Al} shell, placed on bottom gates covered by local \ce{HfO2} dielectric. The \ce{Al} shell had a small segment removed by wet etching, forming the semiconducting weak link of the JJ. Crossovers formed by crosslinked resist on \ce{HfO2} tied the ground plane together across cQED features. NWs and crossovers were contacted to the \ce{NbN} film by ion milling followed by sputtering of \ce{NbTiN}. The sample was wire-bonded and placed in an \ce{In}-sealed \ce{CuBe} box filled with Eccosorb, which was loaded into a dilution refrigerator with a base temperature of ${ \sim \SI{30}{\milli\kelvin} }$. After passing through the sample, microwave signals were amplified by a traveling-wave parametric amplifier at base temperature and a cryogenic amplifier at \SI{4}{\kelvin}. A 6-1-1~T vector magnet was used to apply a magnetic field in the plane of the substrate, along the axis of the NW.

Measurement of the charge-parity switching rate was first performed at base temperature and zero magnetic field. In order for the dispersive QPP signal to be visible, the qubit frequency ${ f_{01} }$ must be in the vicinity of the readout resonator frequency ${ f _{\textrm{RO}} = \SI{5.4}{\giga\hertz} }$. Setting ${ V_C = \SI{-0.71}{\volt} }$ gave ${ f_{01} = \SI{4.1}{\giga\hertz}}$, detected by two-tone spectroscopy. The transmission at the resonator frequency was monitored for ${ \SI{3.2}{\second} }$, yielding time traces as shown in Fig.~\ref{fig:fig1}(c). The aggregated power spectral density ${ P _{xx} }$ of ${ 40 }$ binned time traces was fit using a Lorentzian form
\begin{equation}
  S(f) = A \frac{T_P}{(\pi T_P f) ^2 + 1} + B \, ,
\end{equation}
with poisoning time ${ T_P }$, amplitude $A$ and background $B$ as fit parameters. Experimental data along with the resulting fit are shown in Fig.~\ref{fig:fig1}(d), yielding a low-temperature zero-field poisoning time of ${ T_P \sim \SI{0.1}{\second}}$.  This is on the higher end of values reported in similar systems.

\begin{figure}
\includegraphics[width=8cm]{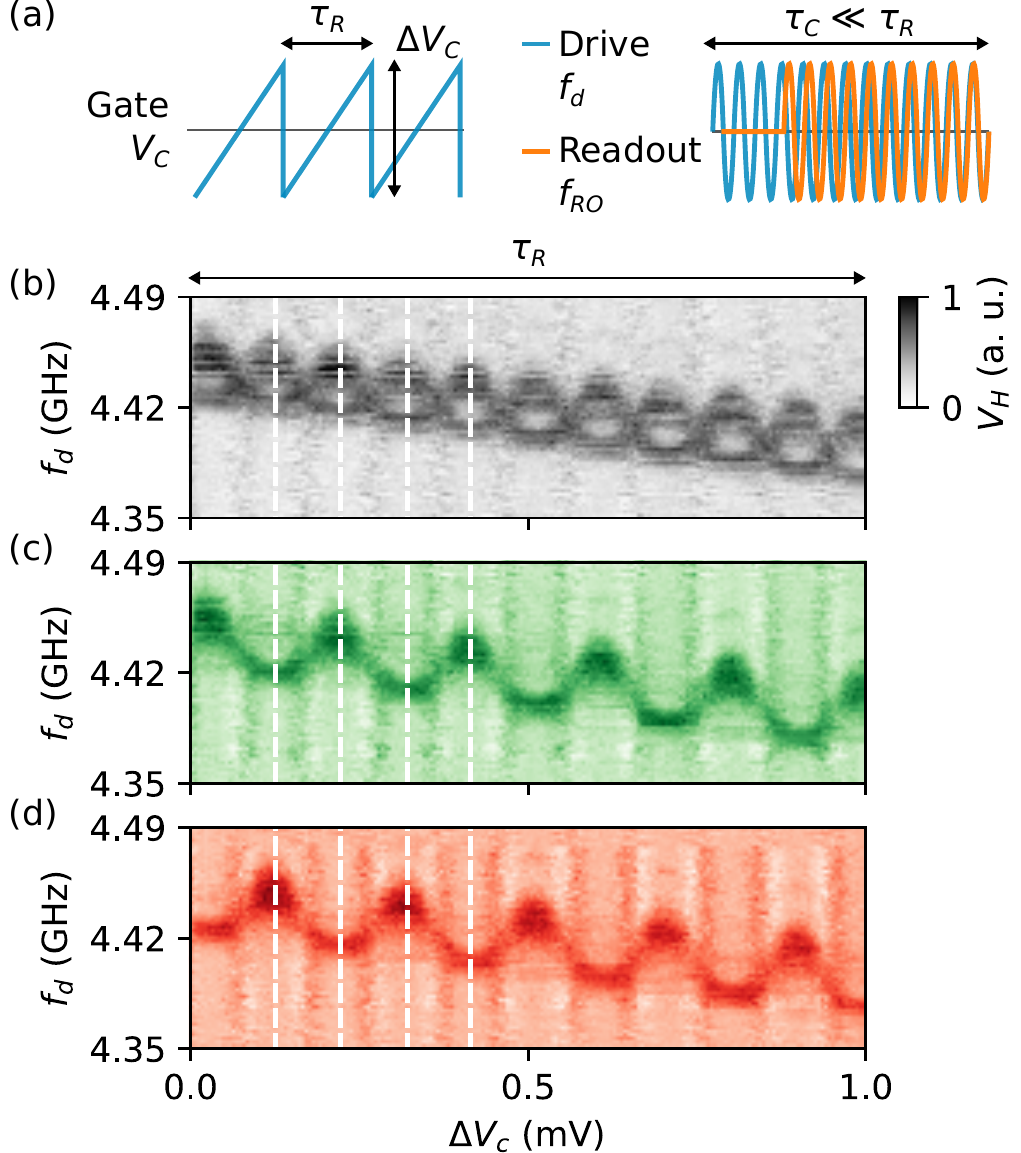}
\caption{\label{fig:fig2} Parity-resolved spectroscopy of transmon charge dispersion. (a) Gate voltage and microwave pulse scheme for measurement. ${ V_C }$ is ramped, while drive and readout tones are transmitted through the device. (b) All data averages, showing two peaks at each gate value (even and odd charge-parity branches). (c-d) Due to slow quasiparticle poisoning compared to the duration of the spectroscopy measurement, the data can be post-selected as into distinct parity branches. The small ${ \Delta V _C }$ used here to control ${ n_g }$ also varies ${ E_J }$, resulting in the overall slope of features in (b-d). Vertical white dashed lines are guides to the eye. Column averages are subtracted for enhanced visibility.}
\end{figure}

Large ${ T_P }$ compared to the measurement duration means that the charge-parity branches of the qubit transition spectrum can be probed separately. In order to demonstrate this, the charge dispersion is probed and the data is post-selected according to charge parity. A ramped voltage is applied to the gate ${ V_C }$ and a two-tone spectroscopy measurement is performed, as shown in Fig.~\ref{fig:fig2}(a). The ramped voltage was offset around \SI{-0.33}{\volt}, yielding ${ f _{01} \sim \SI{4.4}{\giga\hertz} }$. Around this offset, a small ramp amplitude ${ \Delta V_C }$ was used to control the effective charge-offset ${ n_g }$. Typically, this results in the pattern in Fig.~\ref{fig:fig2}(b) where both charge-parity branches are visible \cite{schreierSuppressingChargeNoise2008}. However, because the ramp signal and data acquisition can be applied considerably faster than QPP, the two charge-parity branches can be separated by post-selection \cite{supplement}, as shown in Figs.~\ref{fig:fig2}(c,d). A different technique for parity-selective spectroscopy based on interspersed parity measurements was recently demonstrated for Andreev bound states \cite{wesdorp2021dynamical}.

\begin{figure}
\includegraphics[width=8cm]{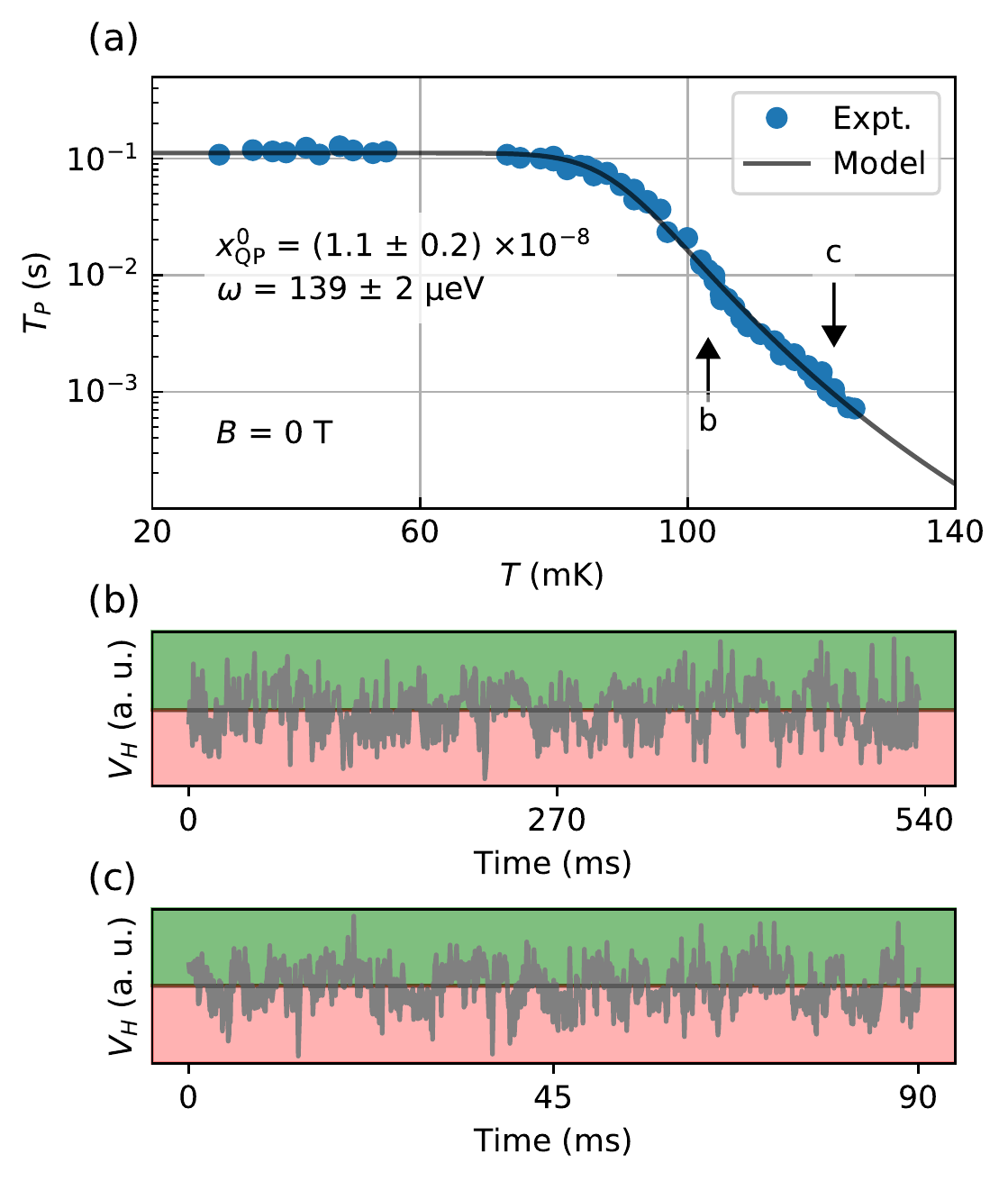}
\caption{\label{fig:fig3} Poisoning time ${ T_P }$ as a function of temperature ${ T }$. (a) As ${ T }$ is increased, ${ T_P }$ is constant then starts to decrease exponentially around \SI{80}{\milli\kelvin}. This is interpreted as the onset of significant contribution to QPP from thermally activated QPs. Model (black curve) yields good decription of experimental data. (b,~c) Examples of individual ${ V_H }$ time traces at different orders of magnitude of ${ T_P }$, as indicated by the arrows in (a).}
\end{figure}

Temperature dependence of QPP is shown in Fig.~\ref{fig:fig3}. Below \SI{80}{\milli\kelvin}, QPP time ${ T_P }$ was found to be independent of temperature ${ T }$. Above \SI{80}{\milli\kelvin}, ${ T_P }$ is exponentially suppressed. Assuming a QP density ${ x _{\textrm{QP}} }$ with a fixed contribution ${ x _{\textrm{QP}} ^0 }$ from non-equilibrium QPs as well as a thermally activated contribution, the total QP density, normalized by the Cooper pair density, is given by \cite{catelaniRelaxationFrequencyShifts2011},
\begin{equation}
  \label{eq:tp}
  x _{\textrm{QP}} = x _{\textrm{QP}} ^0 + \sqrt{2 \pi k_B T / \omega}
  \exp \left( - \omega / k_B T \right)
\end{equation}
where ${ \omega }$ is the superconducting spectral gap. Assuming that ${ 1/T_P = C x _{\textrm{QP}} }$ where ${ C }$ is an unknown proportionality constant, and fitting to the logarithm of data points yields the fit in Fig.~\ref{fig:fig3}(a). The resulting ${ \omega = \SI{139\pm2}{\micro\electronvolt} }$ is in reasonable agreement with the bulk superconducting gap of \ce{Al} ${ \sim \SI{200}{\micro\electronvolt} }$. The density of non-equilibrium QPs ${ x _{\textrm{QP}} ^0 \sim 10 ^{-8}}$ is among the lower values reported for similar systems \cite{serniakHotNonequilibriumQuasiparticles2018, uilhoornQuasiparticleTrappingOrbital2021, pan2022engineering}.

\begin{figure}
\includegraphics[width=8cm]{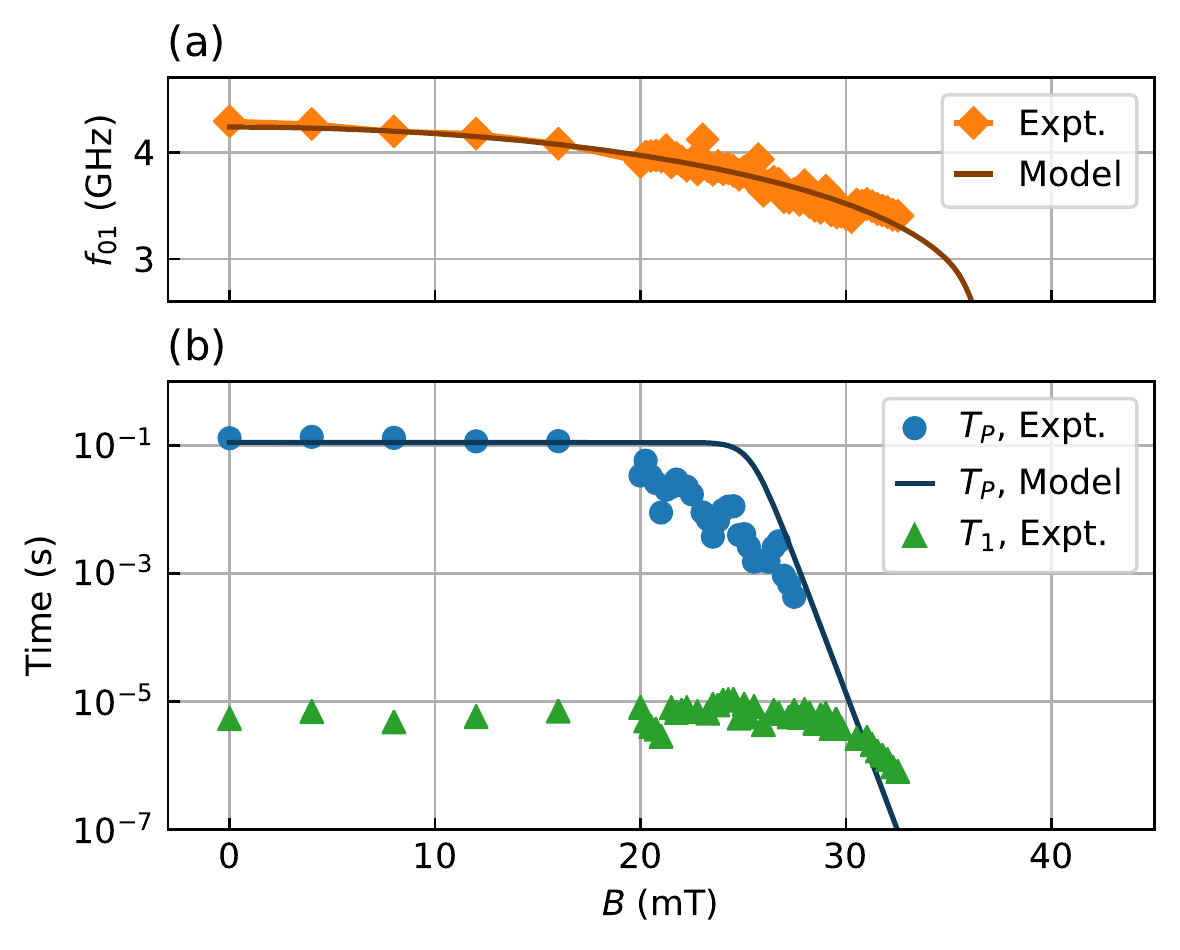}
\caption{\label{fig:fig4} Qubit frequency ${ f_{01} }$, poisoning time ${ T_P }$, and qubit relaxation time ${ T_1 }$ as functions of magnetic field $B$ along the NW axis. (a) ${ f_{01} }$ is suppressed as a function of ${ B }$. The data are fit to ${ \Delta (B) }$ as described by the Little-Parks effect (black curve). (b) ${ T_P }$ is first constant, then diminishes above \SI{20}{\milli\tesla}. As ${ T_P }$ is further suppressed, ${T _1}$ starts to reduce, suggesting that ${T _1}$ becomes limited by QPP.}
\end{figure}

Next, we investigate qubit coherence and QPP as a function of axial magnetic field $B$ relevant for potential applications in topological quantum computation. The full-shell NW exhibits destructive Little-Parks effect \cite{sabonisDestructiveLittleParksEffect2020} which is reflected in the reentrant structure of qubit frequency as a function of $B$ \cite{supplement}. We note, however, that the QP spectral gap ${ \omega }$ is different from the pairing energy ${ \Delta }$, the latter determining the qubit frequency via ${ E_J = (\Delta/4) \sum \eta }$, where ${ \eta }$ are the channel transmissions. Away from zero magnetic field, ${ \omega }$ and ${ \Delta }$ are distinct, as discussed in Sec.~10.2.2 of Ref.~\cite{tinkhamIntroductionSuperconductivity1996a}. At each value of $B$, two-tone spectroscopy was performed and a simple peak-finding algorithm was used to identify the qubit frequency. Rabi and relaxation pulse sequences were then run at the qubit frequency. An exponential fit to the relaxation-sequence data gave ${ T_1 }$, shown in  Fig.~\ref{fig:fig4}.

The ${ f_{01} }$ data in Fig.~\ref{fig:fig4}(a) were fit using ${ f_{01} = f_0 \sqrt{\Delta (B) / \Delta (B = 0)  } }$, where ${ \Delta (B) }$ is governed by the destructive Little-Parks effect due to the full cylindrical superconducting shell of the NW \cite{sabonisDestructiveLittleParksEffect2020, supplement}. The fit yields an NW radius ${ R = \SI{90}{\nano\meter} }$, \ce{Al} shell thickness ${ t = \SI{4}{\nano\meter} }$ and superconducing coherence length ${ \xi = \SI{165}{\nano\meter} }$. The physical dimensions are in rough agreement with transmission electron micrographs of NWs from the same growth batch \cite{supplement}, although ${ t }$ is considerably small. In Fig.~\ref{fig:fig4}(b), measured ${ T_P }$, ${ T_1 }$ are plotted along with a model for ${ T_P }$ as calculated by \eqref{eq:tp} using the parameters from the ${ f _{01} }$ fit, and ${ C }$, ${  \omega (B = 0) }$, ${ x _{\textrm{QP}} ^0 }$ from the fit to temperature-dependent data in Fig.~\ref{fig:fig3}(a) and ${ T = \SI{30}{\milli\kelvin} }$ (base temperature). Here ${ \omega (B) }$ is also governed by the Little-Parks effect \cite{supplement}. The onset of increasing QPP occurs at a lower magnetic field than predicted by the model. The reason for this discrepancy may be related to changes in the quasiparticle density of states at finite field, where it is no longer exponential as assumed in \eqref{eq:tp}. The NW gap is softened at finite field \cite{changHardGapEpitaxial2015} and the coherence peaks are broadened \cite{tinkhamIntroductionSuperconductivity1996a}. Furthermore, the magnetic-field dependence of the non-equilibrium QP population is unknown.

Although QPP imposes a limit on qubit coherence, the ground state switching measured in this work (${ \ket{0e} \leftrightarrow \ket{0o} }$) does not itself relax the qubit, since it is a transition between different charge-parities of the qubit ground state. However, the rate of QP-induced transitions that contribute to relaxation (${ \ket{0e} \leftrightarrow \ket{1o} }$ and ${ \ket{0o} \leftrightarrow \ket{1e} }$) are also related to the QP density, and the similar magnetic fields at which ${ T_P }$ and ${ T_1 }$ start to decrease in  Fig.~\ref{fig:fig4}(b) suggests that QPP becomes the limiting factor on the qubit lifetime above \SI{30}{\milli\tesla}. The full-shell NW exhibits the Little-Parks effect, causing ${ \Delta }$ and ${ \omega }$ to revive as the field is further increased, with a first reentrant lobe centered around \SI{100}{\milli\tesla} \cite{supplement}. At \SI{100}{\milli\tesla}, ${ \omega }$ is predicted to have sufficiently recovered such that ${ T_P }$ is again dominated by non-equilibrium QPs. At this magnetic field, we measured ${ T _{1} = \SI{0.4}{\micro\second} }$, however, no split populations was observed in the dispersive charge-parity monitoring with ${ f _{01} = \SI{3.5}{\giga\hertz} }$, which we interpret as ${ T_P }$ not sufficiently exceeding the measurement integration time ${ \tau _I = \SI{30}{\micro\second} }$. A likely cause of the enhanced QPP in the first lobe is the increased density of subgap states in that regime, as previously observed in the same NWs \cite{kringhojAndreevModesPhase2021}, increasing the QP density of states.

In conclusion, we have examined the rate of charge-parity switching in a full-shell superconducting-semiconducting nanowire qubit. We found a long parity switching time ${ T_P} \sim \SI{0.1}{\second}$ at zero field, and exponential suppression of $T_P$ with field and temperature, in good qualitative agreement with our model. We interpret the exponential onset as associated with thermally activated QPs. Above \SI{20}{\milli\tesla}, this is observed as the rate of QPP rapidly increasing, coinciding with the collapse of the gap in the zeroth lobe of the Little-Parks effect. In the first lobe of reentrant superconductivity, ${ T_P }$ does not recover along with ${ \Delta }$, consistent with previously observed supgap states in this regime. The large value of ${ T_P }$ at zero magnetic field indicates that QPP is not a near-term barrier for the engineering of increased coherence times in this type of qubit.

\section{Methods}
Data in Figs.~1 c-d: A total of ${ N_T = 40 }$ time traces of the in-phase and quadrature components of transmission response are recorded by heterodyne demodulation. Each trace consists of ${ N_P = 800 }$ points separated by ${ \tau _C  = \SI{4}{\milli\second}}$ and integrated over ${ \tau _I  = \SI{100}{\micro\second}}$. For improved signal-to-noise ratio, the data are then projected onto a rotation angle in the phase-quadrature-plane, selected by eye, yielding a scalar ${ V_H }$ \cite{supplement}. A single time trace is shown in Fig.~\ref{fig:fig1}(c). Following the analysis in Ref.~\cite{risteMillisecondChargeparityFluctuations2013}, the data are binned such that points above (below) the mean value are assigned to ${ 1 }$ (${ -1 }$). The binned data are then fit to the Lorentzian form in Eq.~\eqref{eq:tp}.

Data in Figs.~2 c-d: To control ${ n_g }$, a ramp signal is applied to ${ V_C }$ with a period \SI{2}{\milli\second}, amplitude ${ \Delta V_C = \SI{1}{\milli\volt} }$ and offset \SI{-0.33}{\volt}. This value of ${ V_C }$ is sufficiently small so that the gate mainly controls ${ n_g }$ and has a small effect on ${ E_J }$ (visible as the overall slope of features in Fig.~2). This is shown in Fig.~\ref{fig:fig2}(a), along with a variable qubit drive tone ${ f_d }$ and a readout tone at ${ f _{\textrm{RO}} = \SI{5.4}{\giga\hertz} }$. For each value of ${ f_d }$, ${ V_H }$ is measured for ${ 15 }$ consecutive ramps, with ${ 200 }$ data points separated by ${ \tau _C = \SI{10}{\micro\second} }$ taken per ramp period. These ${ 15 }$ ramp periods are averaged into one trace, plotted horizontally in Fig.~\ref{fig:fig2}(b-d). Accordingly, one trace takes ${ 15 \times 200 \times \tau _C = \SI{30}{\milli\second} }$ to acquire. The drive tone ${ f_d }$ is stepped through a window of frequencies, and the entire process is then repeated ${ 20 }$ times. In Fig.~\ref{fig:fig2}(b), all ${ 20 }$ traces are averaged for each ${ f_d }$, yielding a pearl-shaped pattern corresponding to the qubit charge dispersion.

Data in Fig.~3: For ${ T_P }$ to be probed over multiple orders of magnitude, it was not possible to acquire all data at fixed ${ N_T }$ and ${ \tau _C }$. This is because small ${ \tau _C }$ is required to resolve small ${ T_P }$, but such a sampling rate could not be sustained for the duration needed to capture ${ T_P \gg \tau _C }$ due to memory limitations in data acquisition. Therefore, we repeated the temperature sweep trying different combinations of ${ N_T }$, ${ \tau _C }$ and ${ \tau _I }$ at different ${ T }$ until we could measure ${ T_P }$ over a wide range of ${ T }$. Furthermore, for simplicity, only integration times ${ \tau _I > T_1 }$ are used, in order to predominantly detect the ground states. Due to residual excited state population, all four states in Fig.~\ref{fig:fig1}(a) would be resolved for lower ${ \tau _I }$, and the transition rates would need distinguishing. For ${ \tau _I > T_1 }$, the visibility of the ground states dominates, and the analysis is straightforward. At base temperature and ${ B = 0 }$, ${ T_1 \sim \SI{5}{\micro\second} }$ so we set ${ \tau _I > \SI{20}{\micro\second} }$ for all data in Fig.~3 and Fig.~4. This simplification defines the lower bound of detectable ${ T_P }$ in those measurements (${ T_P \gg \SI{20}{\micro\second} }$).

Data in Fig.~4: Measurements of ${ T_P }$ are interleaved with measurements of ${ f_{01} }$ and ${ T_1 }$. To achieve this, a measurement sequence was set up as follows. At each setpoint of ${ B }$, two-tone spectroscopy  was performed from which the qubit frequency was determined by a peak-finding routine. At this frequency, standard Rabi and relaxation pulse schemes were used to determine the $ \pi $ pulse duration and ${ T_1 }$, respectively. Finally, a measurement of ${ T_P }$ was performed; as for the data in Fig.~3, different combinations of ${ N_T }$, ${ \tau _C }$ and ${ \tau _I }$ were used to resolve switching events at different ${ B }$. At \SI{104}{\milli\tesla}, where no split population was observed, the measurement was performed using ${ N_T = 5 }$, ${ N_P = 2000 }$, ${ \tau_C = \SI{100}{\micro\second} }$ and ${ \tau_I = \SI{30}{\micro\second} }$.

\section{Acknowledgments}
We thank Arno Bargerbos, Gijs de Lange, Leonid Glazman, Torsten Karzig, Dmitry Pikulin, Willemijn Uilhoorn, and Bernard van Heck  for useful discussions. We thank Will Oliver for providing the traveling-wave parametric amplifier used in the experiment. We thank Martin Espiñeira for electron microscopy. We thank Marina Hesselberg, Karthik Jambunathan, Robert McNeil, Karolis Parfeniukas, Agnieszka Telecka, Shivendra Upadhyay, and Sachin Yadav at QDev, and Mahesh Kumar, Rizwan Ali, Tommi Riekkinen, and Pasi Kostamo at Espoo for device nanofabrication. Research is supported by the Danish National Research Foundation, Microsoft, European Research Commission grant 716655, and a grant (Project 43951) from VILLUM FONDEN.

\bibliography{aipsamp}

\newcommand{\beginsupplement}{%
     
        \setcounter{figure}{0}
        \renewcommand{\thefigure}{S\arabic{figure}}%
     }
\clearpage
\onecolumngrid
\section{Supplementary Material}
\maketitle
\beginsupplement

\subsection{Rotation of data in phase-quadrature-plane}

\begin{figure}[!h]
\includegraphics[width=18cm]{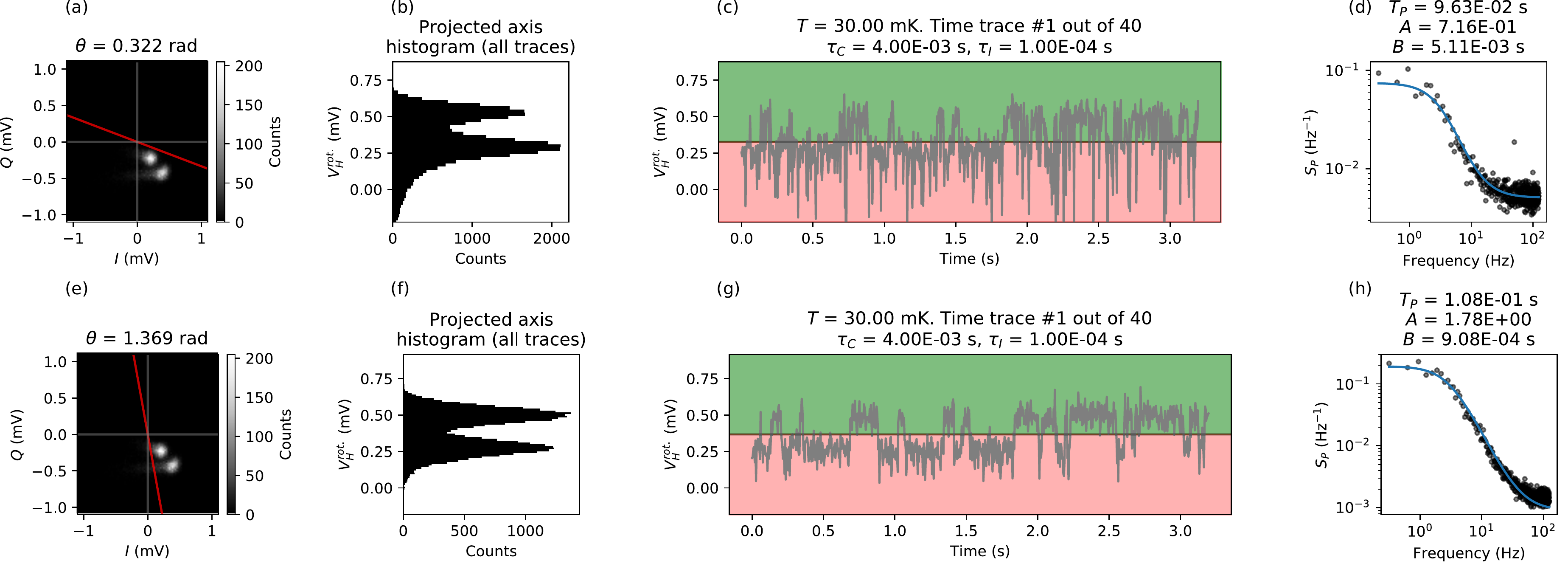}
\caption{\label{fig:suppl-rotation} Plots used to select phase-quadrature-plane rotation angle ${ \theta }$. Two different ${ \theta }$ for the data set in Fig.~\ref{fig:fig1}(c-d). In (a-d) ${ \theta = \SI{0.322}{\radian} }$. (a)~Two-dimensional histogram with the red line indicating ${ \theta }$. The extension of each population are interpreted to result from residual excited state population. (b)~One-dimensional histogram of the data rotated by ${ \theta }$. (c) One of the ${ N_T }$ traces; here ${ N_T = 40 }$. (d) Average PSD of all ${ N_T }$ data sets. (e-h) as (a-d) but with ${ \theta = \SI{1.369}{\radian} }$, which was selected for obtaining the ${ T_P }$ value in the main text from this data.}
\end{figure}

Here we describe the procedure for determining a rotation angle ${ \theta }$ in the phase-quadrature-plane. For each measurement of ${ T_P }$ in the main text, a series of plots were generated for several different ${ \theta }$, as shown in Fig.~\ref{fig:suppl-rotation} for the data set in Fig. 1(c-d) of the main text. The value of ${ \theta }$ was selected to emphasize the bimodality of the two charge-parity populations, as visually apparent from histograms and time traces. Due to instability of ${ n_g }$ over time, the separation of the charge-parity population varies. Data sets where two main populations were not observed were not used to obtain values of ${ T_P }$; this situation could result e.g. from ${ n_g }$ on a degeneracy point, or a charge jump during the course of the measurement. For many data sets, phase-quadrature-plane histogram contained tail-like features, interpreted as partially resolved residual excited state population. In these cases, ${ \theta }$ and angle was selected in a compromise between the axis connecting the charge-parity populations and the line orthogonal to the one connecting the residual features. In general, this is not the same as the line along the midpoints of the two populations. In the case of Fig.~\ref{fig:suppl-rotation}, this meant that ${ \theta = \SI{1.369}{\radian} }$ was selected (bottom row of plots).
The particular criterion used in the selection of ${ \theta }$ is not expected to introduce a bias in ${ T_P }$, as the rotation does not directly alter time domain aspects of the data. For example, in the case of the data in Fig. 1(c-d) of the main text, the standard deviation of ${ T_P }$ values resulting from the Lorentzian fit, across a \SI{1}{\radian} interval around  the selected angle, was found to be ${ < 5\% }$.

\subsection{Parity data post-selection}

\begin{figure}[!h]
\includegraphics[width=8cm]{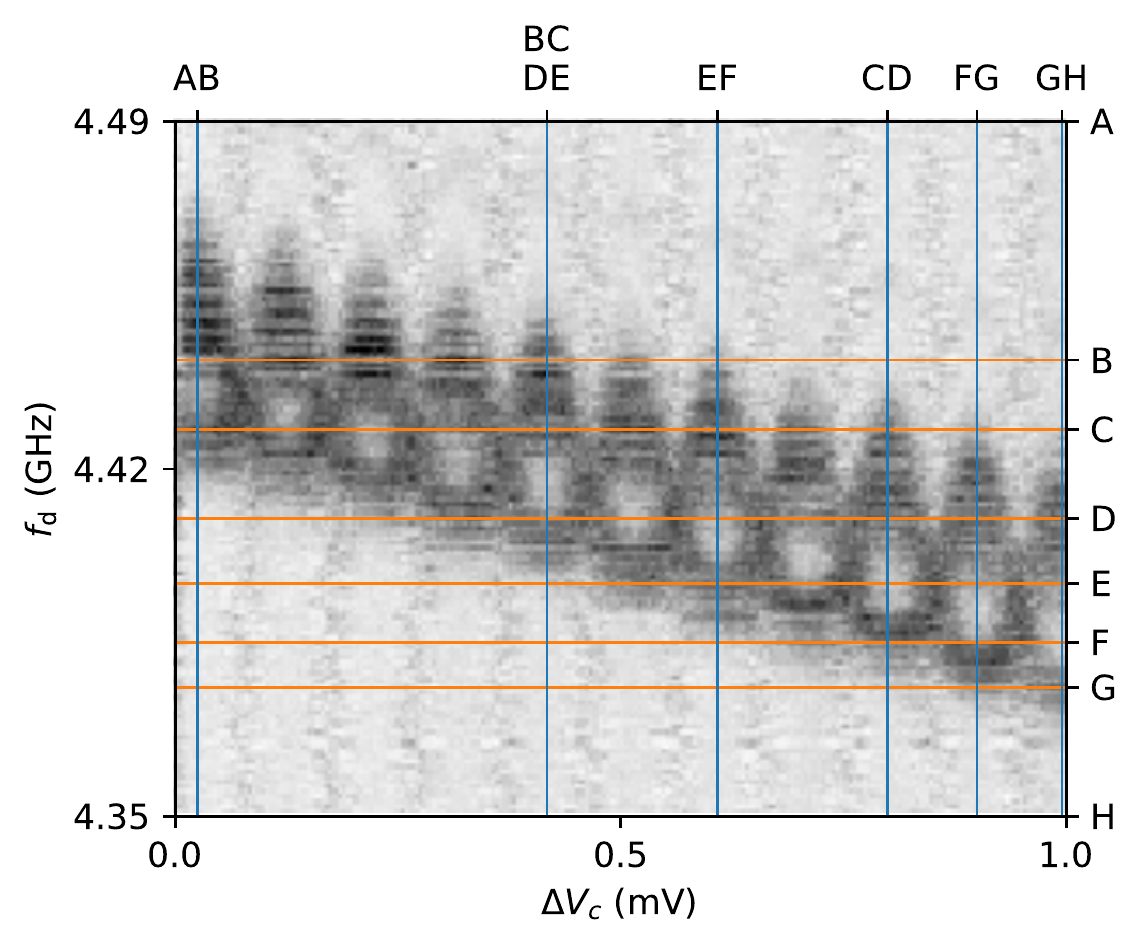}
\caption{\label{fig:suppl-postselection} Criterion points used in the parity post-selection for spectroscopy. For a given interval of ${ f_d }$, data are post-selected conditioned on the value at a chosen ${ \Delta V _c }$ (e.g. at AB for $A > f_d > B$ ) exceeding the mean of that trace of ${ V_H }$ as a function of ${ V_c }$.}
\end{figure}

Here we describe how the two-tone spectroscopy data in Fig.~\ref{fig:fig2}(b) were post-selected into separate parity branches. As shown in Fig.~\ref{fig:suppl-postselection}, the rows are divided into segments AB, BC, etc. Rows are post-selected into parities based on whether the value at a specific point is higher or lower than the mean of that row. The columns picked for the sorting are shown as veritcal lines in Fig.~\ref{fig:suppl-postselection}. This particular procedure relies on a feature unique to one of the parities existing in every row.

\subsection{Destructive Little-Parks effect}

\begin{figure}[!b]
\includegraphics[width=8cm]{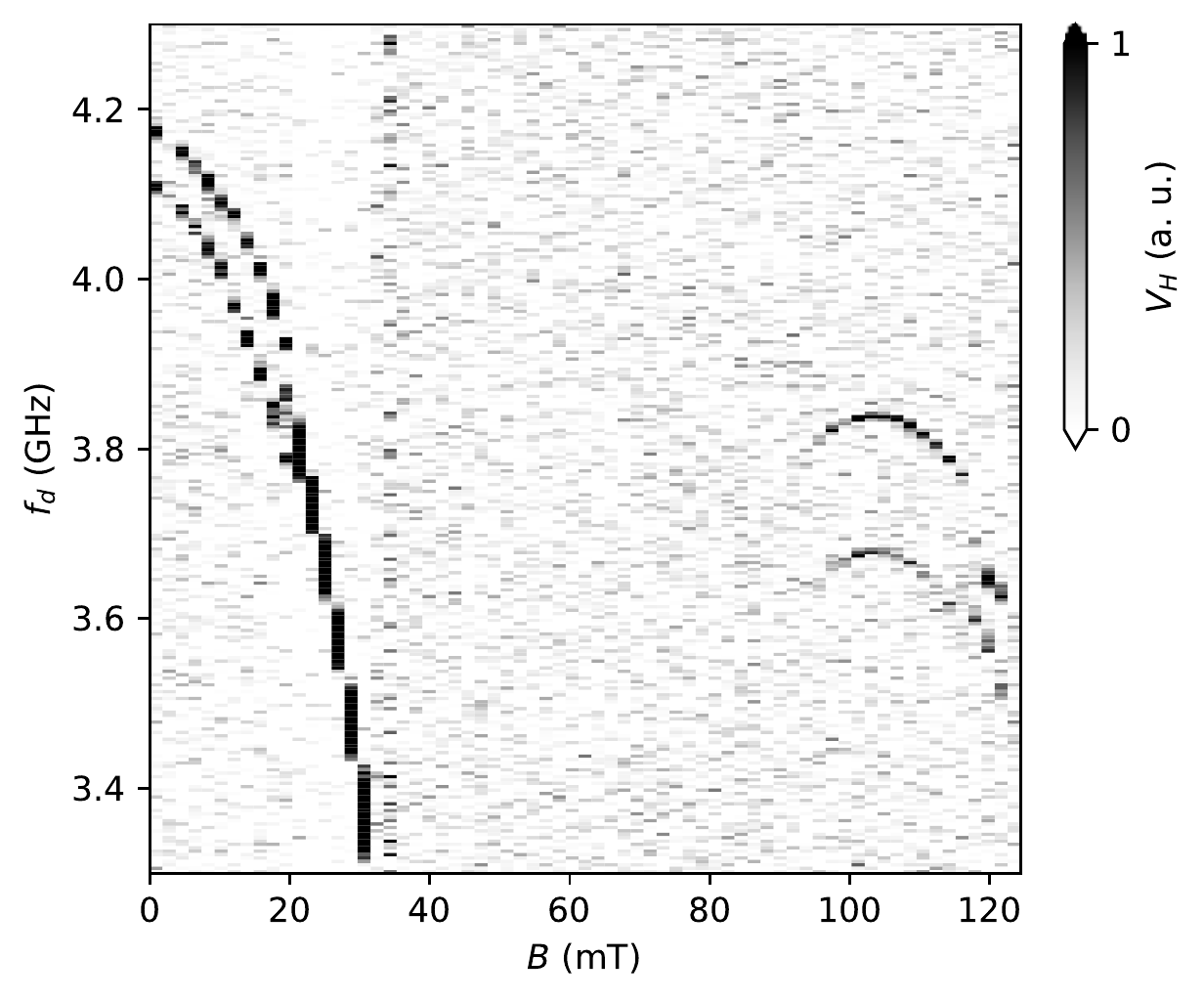}
\caption{\label{fig:suppl-lp} Destructive Little-Parks effect observed by measuring the qubit frequency ${f _{01} }$, resulting in a lobe structure as a function of magnetic field ${ B }$.}
\end{figure}

Similar to Ref.~\citenum{sabonisDestructiveLittleParksEffect2020}, the NW for which we measure ${ T_P }$ here exhibits a destructive Little-Parks effect. This can be directly seen in the two-tone spectroscopy measurement of qubit frequency ${ f_{01} }$ as a function of magnetic field, shown in Fig.~\ref{fig:suppl-lp}. Following Ref.~\citenum{sabonisDestructiveLittleParksEffect2020}, the pair-breaking term ${ \alpha }$ is given minimizing the function \cite{sternfeldMagnetoresistanceOscillationsSuperconducting2011, vaitiekenasAnomalousMetallicPhase2020, daoDestructionGlobalCoherence2009},
\begin{equation}
  \alpha (B) = \frac{4 \xi ^2 k_B T _{C0}}{\pi R^2}
  \left[  \left( n - \frac{\Phi}{\Phi_0} \right)^2 +
    \frac{t^2}{4R^2} \left( \frac{\Phi^2}{\Phi_0^2}+ \frac{n^2}{3} \right)\right]
\end{equation}
in the winding number ${ n }$ at each ${ B }$, where ${ \xi }$ is the superconducting coherence length at zero field, ${ T _{C0} }$ is the critical temperature at zero field, ${ \Phi }$ is the applied magnetic flux, ${ R }$ is the radius of the superconducing shell, and ${ t }$ is the shell thickness. The pairing energy ${ \Delta }$ is then found by solving the implicit equation \cite{larkinSuperconductorSmallDimensions1965}
\begin{equation}
  \log \, \frac{\Delta _0}{\Delta} =\left\{
    \begin{array}{ll}
      \frac{\pi \alpha}{4 \Delta}, & \mbox{if $\alpha < \Delta$},\\
      \log \left( \frac{\alpha + \sqrt{\alpha ^2 - \Delta ^2}}{\Delta} \right) - \frac{\sqrt{\alpha ^2 - \Delta^2}}{2 \alpha} +  \frac{\alpha}{2 \Delta} \arcsin \frac{\Delta}{\alpha}, & \mbox{if ${ \Delta > \alpha }$}.
    \end{array}
  \right.  
\end{equation}
We assume that ${ \Delta }$ enters the transmon Hamiltonian as ${ E_J = (\Delta/4) \sum \eta }$ where ${ \eta }$ are the transmissions of the JJ channels. In terms of ${ \Delta }$, the spectral gap ${ \omega }$ is given by \cite{larkinSuperconductorSmallDimensions1965}
\begin{equation}
\omega = \left( \Delta ^{2/3} - \alpha ^{2/3} \right) ^{3/2} \, .
\end{equation}
We assume that ${ \omega }$ enters into Eq.~\eqref{eq:tp}.

\subsection{Transmission electron micrographs of nanowires}

\begin{figure}[!h]
\includegraphics[width=14cm]{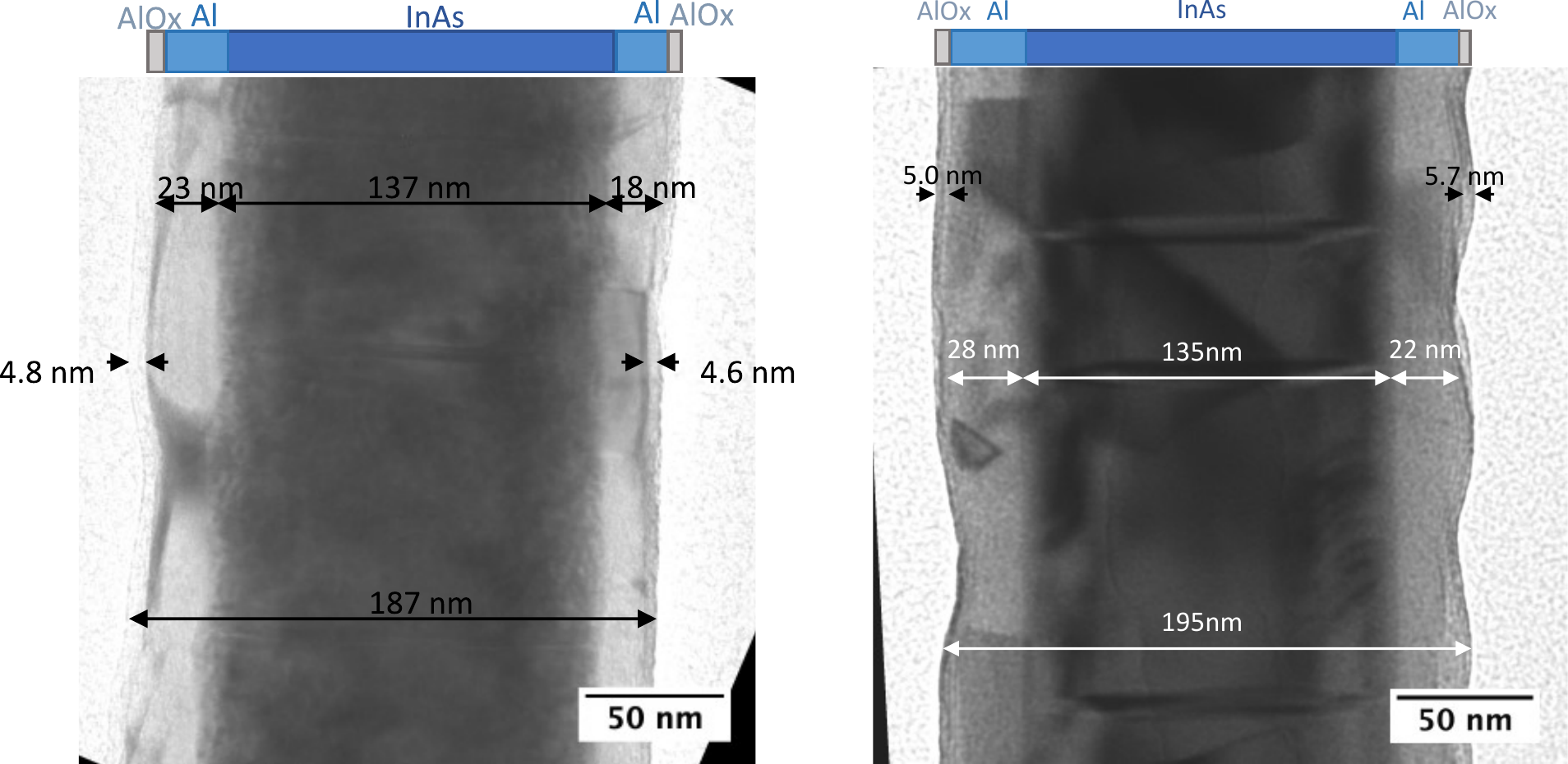}
\caption{\label{fig:suppl-tem} Transmission electron micrographs of two NWs from the same growth batch as the NW discussed in the main text, with physical dimensions indicated.}
\end{figure}

Transmission electron micrographs of NWs from the same growth batch as the NW used in the device described in the main text, are shown in Fig.~\ref{fig:suppl-tem}. Differences between NWs in micrographs and the one used for the device might exist, due to aging, position on wafer, and individual wire-to-wire variations.

\subsection{Experimental setup}

\begin{figure}[!b]
\includegraphics[width=16cm]{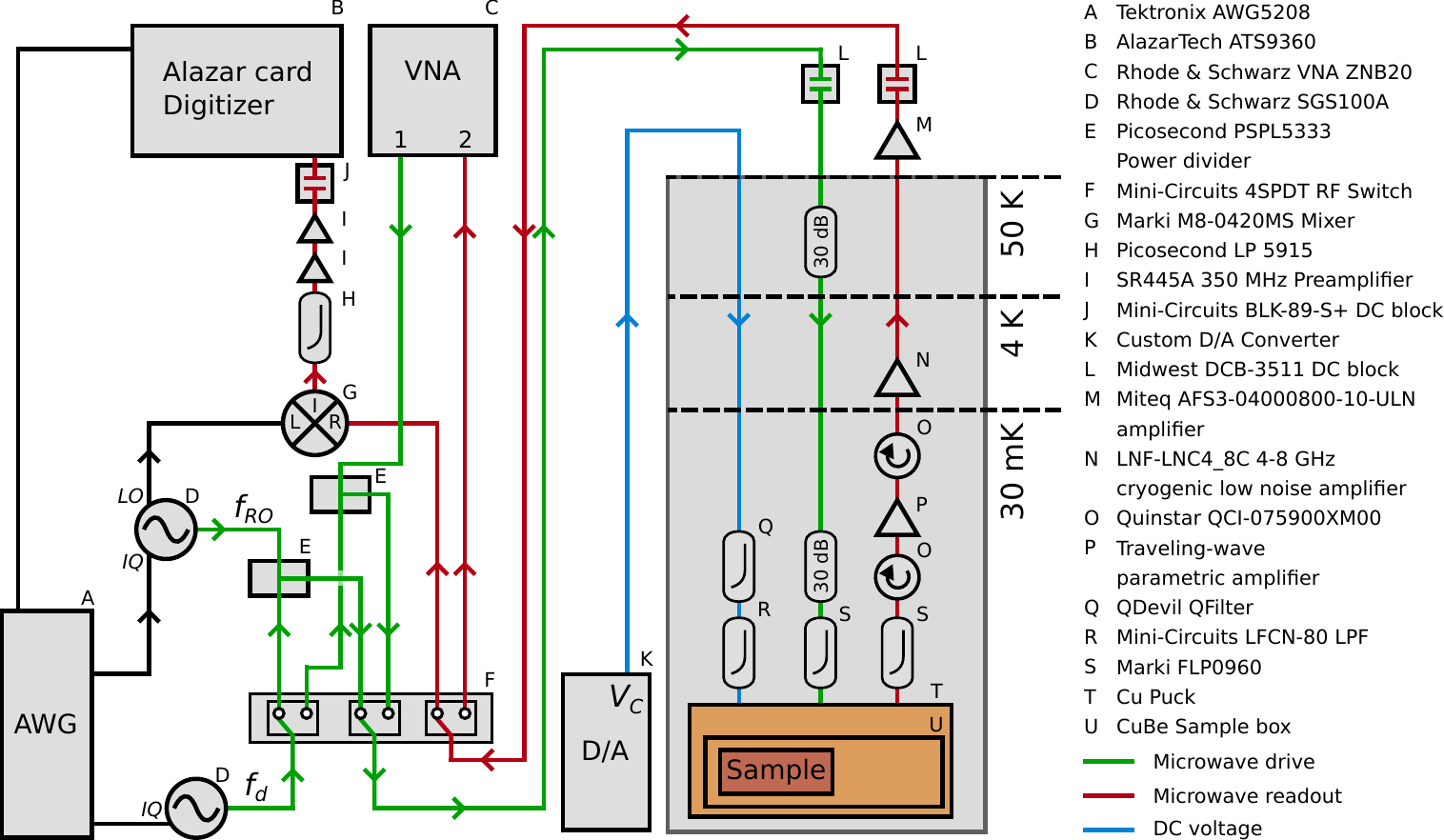}
\caption{\label{fig:suppl-setup} Experimental setup.}
\end{figure}

A diagram of the experimental setup is shown in Fig.~\ref{fig:suppl-setup}. The sample is mounted on a printed circuit board in an \ce{In}-sealed CuBe box, with added Eccosorb foam.

\end{document}